# Evidence for abundant transcription of non-coding regions in the *Saccharomyces cerevisiae* genome

Moshe Havilio[1], Erez Y. Levanon[1,2], Galia Lerman[3], Martin Kupiec[3] and Eli Eisenberg[1,4].

[1] *Compugen Ltd., 72 Pinchas Rosen St., Tel-Aviv 69512, Israel*

[2]*Department of Pediatric Hematology-Oncology, Chaim Sheba Medical Center and Sackler School of Medicine, Tel Aviv University, Tel Aviv 52621, Israel*

[3]*Department of Molecular Microbiology and Biotechnology, Tel Aviv University, Ramat Aviv 69978, Israel*

[4]*School of Physics and Astronomy, Raymond and Beverly Sackler Faculty of Exact Sciences, Tel Aviv University, Tel Aviv 69978, Israel*

Moshe Havilio - moshe@compugen.co.il

Erez Y. Levanon - erez@compugen.co.il

Galia Lerman - galilerman@hotmail.com

Martin Kupiec - martin@post.tau.ac.il

Eli Eisenberg – elieis@post.tau.ac.il

Correspondence: Moshe Havilio E-mail: moshe@compugen.co.il

# Abstract

**Background:** Recent studies in a growing number of organisms have yielded accumulating evidence that a significant portion of the non-coding region in the genome is transcribed. We address this issue in the yeast *Saccharomyces cerevisiae*.

**Results:** Taking into account the absence of a significantly large yeast EST database, we use microarray expression data collected for genomic regions erroneously believed to be coding to study the expression pattern of non-coding regions in the *Saccharomyces cerevisiae* genome. We find that at least 164 out of 589 (28%) such regions are expressed under specific biological conditions. In particular, looking at the probes that are located opposing other known genes at the same genomic locus, we find that 88 out of 341 (26%) of these genes support antisense transcription. The expression patterns of these antisense genes are positively correlated. We validate these results using RT-PCR on a sample of 6 non-coding transcripts.

**Conclusions:** 1. The yeast genome is transcribed on a scale larger than previously assumed. 2. Correlated transcription of antisense genes is abundant in the yeast genome. 3. Antisense genes in yeast are non-coding.

**Background**

Recent systematic searches for transcribed regions have yielded growing evidence suggesting that the fraction of the genome being transcribed is much larger than previously thought [1-4]. In particular, it is becoming clear that a significant fraction of the expressed DNA comes from non-coding regions [2]. The study of the role and function of this large amount of non-coding RNAs attracts much attention and effort. The yeast *Saccharomyces cerevisiae* is one of the most studied model organisms, and the first eukaryote to have its genome sequenced [5]. Its genomic structure is by far simpler than that of mammalian cells, as splicing and alternative splicing play only a minor role. The early availability and simple structure of the yeast genome made it unnecessary to apply transcriptome-based methods such as large-scale sequencing of ESTs. Instead, it was first assumed that as a good first approximation, its transcriptome was well represented by the set of all open reading frames (ORFs) longer than 100 amino acids. This set of ORFs accounts for most, if not all, coding genes, but ignores the non-coding expressed regions. As a result, although much high-throughput expression data have been accumulated for the yeast ORFs, far less is known on the expression of non-coding regions.

Recent studies compared several yeast genomes and showed that many of the yeast ORFs believed to be protein-coding genes are actually not conserved even in closely related species [6-9]. These studies suggested that the non-conserved ORFs do not actually code for proteins. These ORFs (825 out of 6703 yeast ORFs; 12%) were thus called "dubious ORFs" (DOs) [10]. In this work, we use the expression data collected for these DOs to study the expression pattern of non-coding regions in the yeast genome. We rely entirely on the current DO annotation and do not attempt to find new such ORFs. The probes in commercial yeast DNA microarrays (such as those developed by Affymetrix) were chosen to represent all the putative ORFs larger than 100 amino acids, and thus include all the ORFs now classified as dubious. Since the dubious ORFs are now known to be non-conserved and thus presumably non-coding, it follows that one can treat the expression measured by the corresponding probes as representing the whole non-coding parts of the yeast genome. While Affymetrix probes are biased to the 3' end of the ORF, this bias is irrelevant, as we know that no protein is encoded in this region. Our results on the expression levels of dubious ORFs are thus not limited to these ORFs only. One should consider these genomic regions as a random sampling of the whole non-coding part of the yeast genome.

A surprising abundance of antisense transcripts, RNAs transcribed from opposing DNA strands at the same genomic locus, was recently observed in several eukaryotic genomes [4, 11-13]. Some antisense transcripts have been shown to regulate gene expression [14, 15] but in general not much is known about how antisense transcription regulates gene expression in mammalian cells. A large fraction of the DOs show partial overlap with other, usually conserved, ORFs encoded on the opposite strand [16]. In fact, almost all of the sense/antisense (S/AS) ORF pairs in yeast are non-DO/DO pairs: 501 of the 538 pairs (93%) are DO/non-DO, while only 22 pairs are DO/DO, and 15 are non-DO/non-DO respectively. Thus, the extent of the antisense phenomenon in yeast critically depends on the prevalence of DO transcription. In the following, we show that a large fraction of the DOs are actually expressed. The expression pattern of the S/AS pairs is analyzed, showing that the expression profiles of two antisense transcripts are correlated.

## Results

**Expression of dubious ORFs**

We started by accumulating a large body of Affymetrix gene expression experiments, which includes 154 experiments performed in 12 different studies. Background subtraction and normalization procedure for this data set are described in Methods. The dataset includes recorded expression data for 6437 and 589 non-DOs and DOs, respectively. In the following we use this data to show that many of the DOs are actually expressed. Moreover, we find that the expression levels of S/AS DO/non-DO pairs are correlated. This latter result further supports the idea that the expression of the dubious genes is a true biological phenomenon rather than an artifact or a random experimental error.

First, we analyzed the expression profiles of the DOs, in order to see whether some of them are expressed after all. The microarray expression profiles were searched, looking for DOs that were expressed at high levels in particular conditions. We define a DO as a "strongly expressed DO" (SDO) if its level of expression exceeds the $70^{th}$ percentile in at least one condition in our data set, and it does not overlap any non-DO. This expression criterion depends only on the expression intensity of the ORFs relative to other probes in the same experiment, and hence does not depend on the background subtraction and normalization of different data sets. The 70th percentile threshold roughly corresponds to a threshold of 200 in normalized Affymetrix average difference (PM-MM) units, which is a conservative cutoff that minimizes the false positives rate [17, 18]. We find 164 such SDOs (28% of all DOs with recorded expression), which are listed in Supplementary Table 1 along with their 5 highest expression values. These 164 SDOs contain 88 antisense DOs and 76 non-antisense DOs, which are 26% and 31% of the AS and non-AS DO with recorded expression respectively. Hence, the strong expression of the SDOs cannot be accounted for by the antisense phenomenon, nor attributed to an artificial connection between the strands, like in cDNA expression experiments. One thus may conclude that at least 28% of the DOs are actually expressed into RNA under specific biological conditions.

**<u>Antisense expression correlations</u>**

We then studied the relation between the expression profiles of S/AS pairs. 495 DOs participate in DO/non-DO S/AS pairs (6 DOs have two overlapping non-DOs), of which 341 have recorded expression and 333 exceed the background level (see Methods). We compared the distribution of Pearson's correlations (PC) [19] between the expression profiles of these DO/non-DO S/AS pairs with that of three control sets: (i) Randomly picked non-DOs from the S/AS set, paired with randomly picked DOs (see Figure 1) (ii) Randomly picked DOs from the S/AS set, paired with randomly picked non-DOs (iii) Randomly picked DOs paired with randomly picked non-DOs (It is important to use random non-DO/DO pairs as control since the expression level of DOs is, on average, three times lower as that of non-DOs). If the observed expression of DO is a result of random experimental error, or caused in any way by our normalization procedure, we would expect no significant difference between the PC distribution of S/AS DO/non-DO pairs and that of one of the random sets of DO/non-DO. The results are summarized in the first row of Supplementary Table 3. A significant difference is observed ($P<2\times10^{-10}$; $\chi^2$ test) between the S/AS PC distribution and the PC distributions for each one of the other random sets, the S/AS pairs being more correlated.

Employing the "leave one out" method, we further checked that the results do not follow from one single study, which is either faulty, use inadequate experimental technique, or not compatible with our processing methods. We carried out statistical tests leaving each time one of the data sets out of the full data set. Table 3 in the Supplementary Information shows that regardless of the study left out or the control set of random pairs chosen, there is a significant difference between S/AS DO/non-DO and random pairs PC distribution, S/AS pairs being on average more correlated than random pairs.

Thus, our results show an abundant transcription of the strand opposite to coding genes in yeast, where the opposite strand expresses a DO. The expression of the two strands is correlated.

**Experimental confirmation**

In order to experimentally confirm the expression of DOs, we carried out an RT-PCR analysis. First, the expression of 4 SDOs was analyzed. Oligonucleotides specific for ORFs *YER121w, YMR245w, YDR525w* and *YJL199c* were synthesized, and RT-PCR analysis was carried out. In this assay RNA products created by transcription of specific genes are first amplified by reverse-transcription, and the resulting products are further amplified by PCR. An RT- PCR product of the expected size was obtained for each of the four analyzed genes (Figure 2A and data not shown). No RT-PCR product was obtained with samples in which the reverse-transcription step was omitted, indicating that the results obtained represent true amplification of mRNA molecules. The results obtained confirm that these ORFs are transcribed. Direct DNA sequencing of the RT-PCR products confirmed the identity of the amplified sequences. Next, we analyzed S/AS non-DO/DO pairs. These include the ORFs *YBR112c*/*YBR113w*, *YGR181w*/*YGR182c* and *YPL181w*/*YPL182c*. All the DOs present in these pairs are expressed, as demonstrated by the RT-PCR product obtained individually with primers specific for each individual ORF (Figure 2B and data not shown). As before, we observed no RT-PCR product when the reverse-transcription step was omitted, again demonstrating the presence of a transcribed RNA molecule. To rule out potential artifacts related to the presence of additional transcripts emanating from adjacent ORFs, we also performed RT-PCR reactions with primers encompassing both ORFs (p1 and p4 in Figure 2B). No amplification was detected, ruling out the possibility that the detection was due to the presence of a single, joint, transcript. In order to reject the possibility of the DO *YGR182c* being part of a long UTR of its adjacent ORF *YGR183C*, we calculated the expression Pearson correlation between these two ORFs and found it to be rather weak, only 0.26. For comparison, the expression correlation between *YGR183C* and its antisense *YGR181w* is 0.39. In addition, for the other pairs tested the transcripts are divergent, so there cannot be a transcript coming from a nearby gene.

We thus conclude that many of the dubious ORFs are transcribed, some of which as sense/antisense pairs. As explained above, genomic comparisons have demonstrated that these dubious ORFs are not conserved among closely related yeast species, and thus probably do not translate into proteins. If indeed these genomic loci do not code for proteins, nothing distinguishes them from any other genomic region. However, we clearly show here that many of these genomic loci are transcribed, sometimes at very high levels. Therefore, our results suggest that a large fraction of the yeast genome is transcribed, even if it does not encode for conserved proteins. In addition, we have shown that yeast cells express numerous sense/antisesnse transcripts that may play a role in controlling gene expression, in accordance with previous reports for other species genomes [4, 11-13, 20]. Hurowitz and Brown [16] have recently analyzed the lengths of yeast mRNAs using the Virtual Northern method. A categorized list of 820 DOs was included in their study; each DO was classified according to the agreement between its observed and expected mRNA transcript length. Transcript measurements were available for 243 of the DOs. Of these, 192 (79%), exhibited good agreement. We find a high degree of association between the SDO set and the transcribed lists: 163 out of 164 SDOs were categorized; 60 of them were classified as exhibiting good agreement, and 77 had some agreement (P=$10^{-5}$ and $7\times10^{-8}$ accordingly; a hypergeometric probability function - see Methods). These experiments thus further support our claim that the SDOs are actually expressed.

It should be noted that the definition of SDOs is cut-off and dataset dependent. Reducing the strong DO expression threshold from 70 percentile to 50 percentile results in an increase in both SDOs and S/AS pairs: 336 SDOs and 214 S/AS SDO/non-DO pairs (compared with 164 and 88, respectively). The addition of the expression data set from Roth et al. (1998) [21] (4 additional conditions) results in a total of 251 SDOs, of which 149 are paired with a non-DO antisense. We chose to ignore this data set due to the sharp increase in the number of SDOs, which is not reflected in any of the other studies used.

## Discussion

We demonstrated that a significant portion of yeast's DOs is expressed. While we cannot rule out the possibility that some of the DOs do actually code for proteins, there is strong evidence against this possibility [6]. In addition, we compared the list of 825 yeast's DOs with a list of homologous pairs of *Ashbya gossypii* and yeast ORFs [7]. Only one out of 825 members of our DOs list (YJR012C) had a homolog among *Ashbya gossypii* genes. This further supports the view that the DOs do not code for proteins. Combining this finding with our strong evidence for expression of the DOs, it follows that a large fraction of the yeast genome is transcribed but not translated. This finding is in accordance with similar results for a number of other organisms studied [1-4].
The question thus arises: what is the role and function of this transcription? Early models of differential gene expression [22] assumed that most of the genome is transcribed, and that untranslated regions, such as the 3' and 5' ends of the genes, could play a central role in regulation of expression. In accordance, a recent study has studied full-length yeast transcripts, and found that untranslated regions (UTRs) encompass about 300 nucleotides per gene, much longer than previously expected. According to this study, 15% of the yeast genome is transcribed as UTRs [16]. UTRs have important roles in regulating mRNA localization and translation [23].

Therefore, some of the expressed DOs could be actually within the 3` or 5` UTR of an adjacent coding ORF. However, this explanation can account for only part of the expressed dubious ORFs we observed, as the distance of many of the expressed ORFs to their adjacent coding ORF is larger than 300bp. It thus appears that these represent non-coding RNA genes. A growing number of non-coding genes, such as small nuclear RNAs and microRNAs, have been recently observed in many different organisms, in addition to the known tRNAs and rRNAs. These RNA genes are transcribed in a regulated way and, although not coding for proteins, play an important role in regulating the expression of other genes. For example, the yeast *SER3* gene has been recently shown to be regulated by the noncoding intergenic transcript *SRG1* [24]. Computational effort has been made to find such genes employing comparative genomics and domain analysis methods [25]. However, genes of this type have been largely over-looked in all large-scale analyses of the yeast genome that were based on the search for ORFs. It is thus possible that non-coding genes cover a significant part of the yeast genome, and account for the expression of part of the dubious ORFs.

Moreover, an important subset of the non-coding DOs is the set of many antisense RNAs – RNAs expressed from the strand opposite to a coding gene – that may act to regulate the translation of the coding gene on the other strand. Our results show that, similar to other eukaryotes genomes [4, 11-13, 20], yeast cells express an abundance of antisense transcripts, and virtually all of these transcripts are non-coding. In addition, we found for the first time an association between the expression levels of sense and antisense genes. These results open the way to further understanding of the antisense phenomena in yeast, which is the most studied eukaryotic model organism.

**Conclusions**

We have provided evidence that a significant portion of the non-coding regions in the yeast genome is transcribed, in accordance with similar results recently obtained for many other organisms. We have shown that the antisense phenomenon in yeast is almost solely limited to non-coding regions, and that correlated transcription of antisense non-coding genes is abundant in the yeast genome. A number of possible regulatory mechanisms based on non-coding transcripts have been suggested, but the overall role of this phenomenon is yet elusive.

**Methods**

Yeast genes: a list of yeast genes was retrieved from Sacchromyces Genome Database [10], ftp://ftp.yeastgenome.org/yeast/data_download/chromosomal_feature/intron_exon.tab). There are 7156 genes on the list; of them 6703 are ORFs, from which 825 are DO. All yeast data is correct to November 2003. We will limit our discussion to the ORFs.

Expression data set: Affymetrix expression data sets were collected from 12 sources [26-37]. Altogether 154 different conditions were assembled. 6437 and 589 non-DO and DO respectively have recorded expression data in our data set.

Background subtraction and normalization: The average background (BG) level for each condition was presumed to be at the conservative level of the twentieth percentile. For comparison, the parallel value used in [38] is five percentile. To avoid BG fluctuations effects, the background level was subtracted from the expression measure, and any expression level below this cutoff was set to 0. This subtraction considerably reduces artificial correlation between weakly expressed ORFs. After BG subtraction, expression levels were converted into relative RNA abundance.

Hypergeometric distribution function [39]: The probability of drawing m+n DOs, containing at least m transcribed DO, out of a categorized list containing M+N transcribed and untranscribed DO respectively is:

$$\Pr(k \geq m \mid n + m) = \sum_{k=m}^{\min(M,m+n)} \frac{\binom{M}{k}\binom{N}{m+n-k}}{\binom{M+N}{m+n}}$$

RT-PCR: RNA was extracted from $5 \times 10^7$ cells using the RNeasy kit (QIAGEN Inc.) according to the manufacturer instructions. Prior to the RT-PCR step, genomic DNA was degraded by RQ1 RNase-free DNase (Promega Inc.). Complete removal of contaminating DNA was verified by negative control PCR reactions with a specific set of primers. 1 µg of total RNA was used as a template for cDNA synthesized using the Expand$^{TM}$ Reverse Transcriptase Kit (Boehringer Mannheim) and 500 ng of oligo (dT)$_{15}$ (Boehringer Mannheim) as a primer. One quarter of each cDNA preparation was used as a template in a PCR reaction using specific primers. The products were subjected to agarose electrophoresis.

Primers: The following primers were used (F; forward, R: reverse):

*YER121w*: CAAGGCCAGCAGAGGAAAAG (F) and ATGTGCGTATGAAGCGGTTG (R).
*YMR245w*: TCTGTATATTCTGTATCTATGTTCCTGC (F) and AAATGGCCTATTGTATTGTCAGGTC (R).
*YDR525w*: CAAGAATTTCTCGAGTTCCTTATATATGAG (F) and AGTTTATTTCCAAAATAGCGAAGACC (R).
*YJL199c*: CGACTGCCGCTGTTCATTCT (F) and CTTCTTGTTGCCGGCCTG (R).

Sense/antisense pairs:
*YGR181w*: P1: CTATCATCTATCTTTGGCGGCG, P2: GGTTGTCTTTGCAGTAGTGGCTG.
*YGR182c*: P3: CAAAGGATACCAAGAAAATGCTATTACG, P4: TGGAAATAGACAGAACGAGCC

*YBR113w*: P1: GTTCACCGCCCGGATTC, P2: TTTTACAAACCACGTCAGGAGTTC
*YBR112c*: P3: ACTGAAGAGGCGGAGCCAG, P4: CAAAGTAGGTTTGATTACAGTTATCGTTG

*YPL181w*: P1: AGTCTGATCGAGAGGAATTTGTACG, P2: CTCTAGTCAGGTCGTCCATCATTG
*YPL182c*: P3: GACCATCAATAGTTTGTTTCCTTCG, P4: CTCTAGTCAGGTCGTCCATCATTG

**Authors' contributions**

MH and EE conceived and designed the research plan and participated in all aspects of data collection and analysis. EYL participated in data analysis and interpretation. GL carried out the molecular studies. MK participated in the design of the study.

**Figure legends:**

**Figure 1:** Distributions of PC for S/AS DO/non-DO pairs (blue) and randomly reshuffled DO/non-DO pairs from the S/AS set (green). 103,479 pairs used to calculate the second distribution. There is a significant difference between the S/AS and random distribution ($\chi^2$=124, DF = 9 and $P = 10^{-22}$). S/AS pairs are significantly more correlated than random DO/non-DO pairs. For the S/AS distribution average= 0.076 and SD=0.22. For the reshuffled distribution average= –0.018 and SD=0.18. The significance of the difference in the averages is $P_t=10^{-20}$ (student t-test).

**Figure 2:** RT-PCR analysis of dubious ORFs. A) Reverse-transcription was carried out using oligo dT and specific primers for each ORF. RT: reverse transcription followed by PCR amplification. N: No reverse transcriptase added. G: Genomic DNA (positive control). NS: No RNA or DNA substrate added (negative control). B) A similar RT-PCR analysis was carried out for sense-antisense pairs. Shown here are YGR181w/*TIM13* and the DO YGR182c. No amplification was detected with primers p1 and p4. If a transcript was present that encompasses both ORFs, amplification was expected, of the size observed in the reaction carried out with genomic DNA

# Additional Material file

File name: Additional_Material.pdf

**Title and description of data:**

*a. Supplementary tables 1+2.*

**Title:** A list of highly expressed non-coding genomic regions.

Description: 164 strongly expressed dubious ORFs along with their 5 highest expression values and the study in which each value is originated.

*b. Supplementary tables 3.*

**Title:** A "leave one experiment out" verification of the sense-antisense expression correlation.

Description: Comparison between Pearson correlation coefficients distributions for S/AS DO-non-DO pairs and various sets of random pairs, while leaving out each of the experimental data sets. This table shows that regardless of the set of random pairs considered or the left out data set sense/anti-sense ORF pairs are significantly more correlated than random pairs.

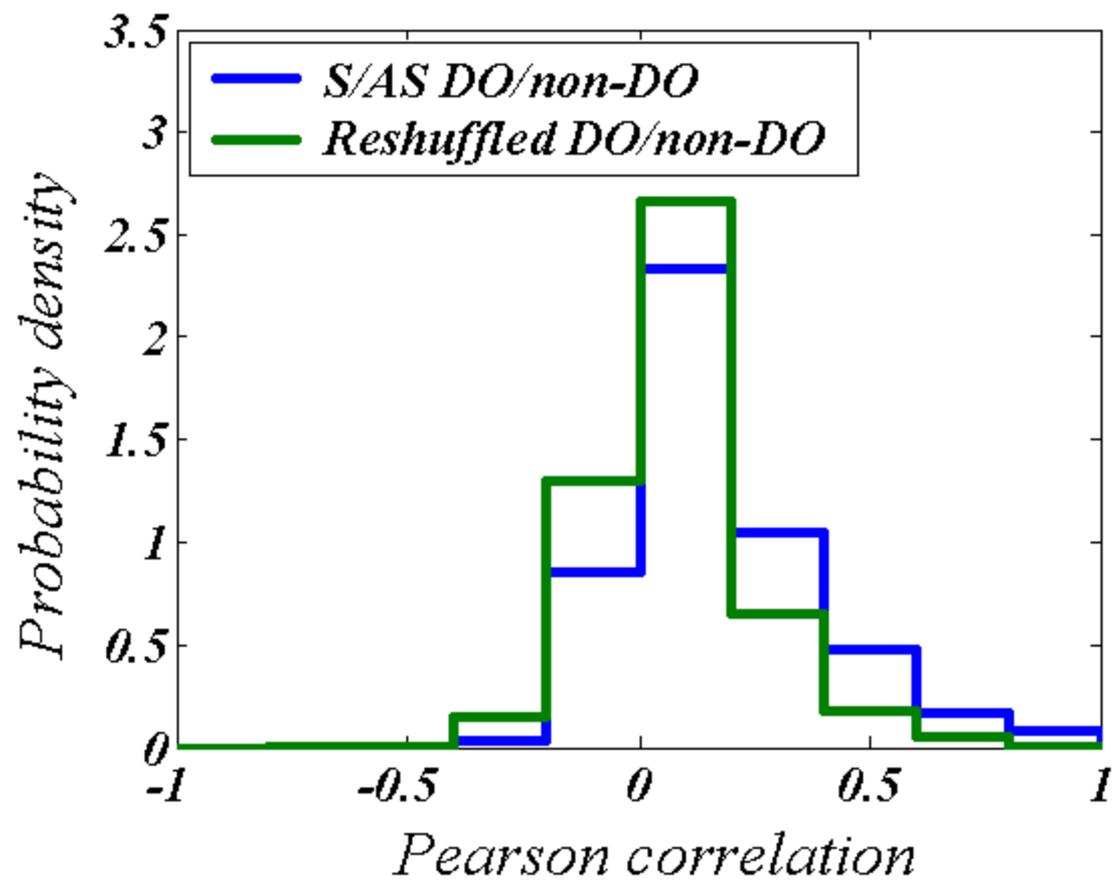

Figure 1

# A

YER121w    YMR245w    YDR525w

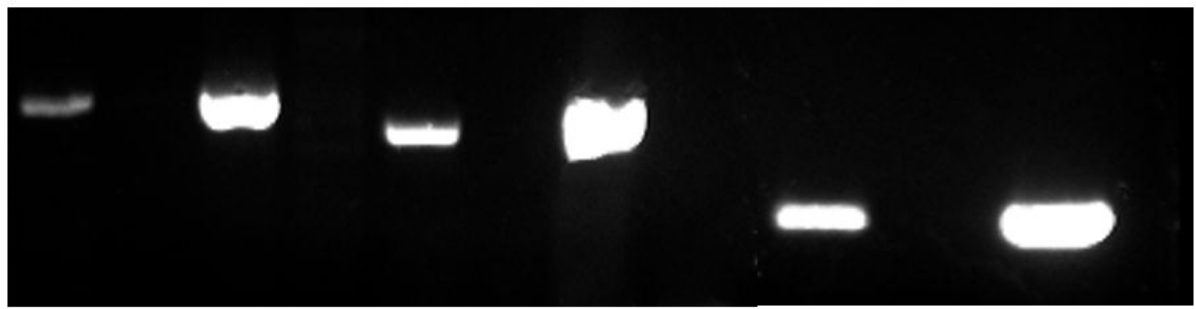

RT N G NS    RT N G NS    RT N G NS

# B

YGR181w (p1+p2)   YGR182c (p3+p4)   (p1+p4)

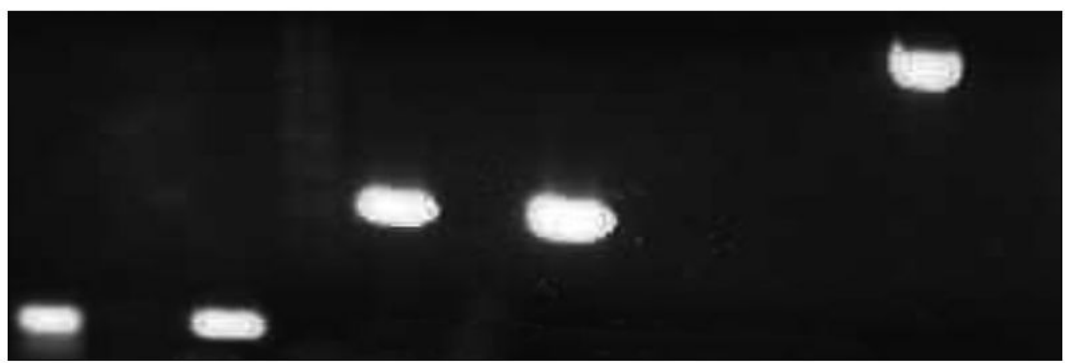

RT N G NS   RT N G NS   RT N G NS

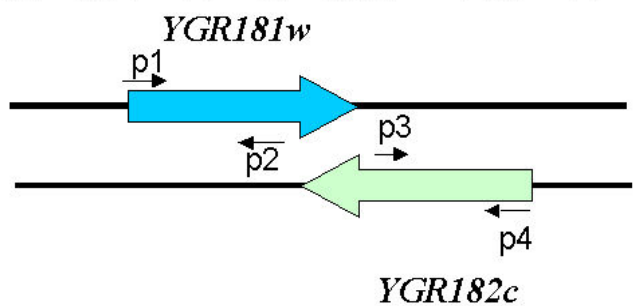

Figure 2

**Additional files provided with this submission:**

Additional file 1 : Additional_Material.doc : 221Kb
http://www.biomedcentral.com/imedia/1637076016717632/sup1.DOC